\input epsf.tex
\documentstyle[12pt]{article}

\newcommand{\be}{\begin{equation}}   \newcommand{\ee}{\end{equation}}
\newcommand{\bear}{\begin{eqnarray}}
\newcommand{\eear}{\end{eqnarray}}
\newcommand{\ba}{\begin{array}}      \newcommand{\ea}{\end{array}}

\newcommand{\ie}{{\it i.e.\ }}

\begin{document}
\pagestyle{empty}
\begin{titlepage}

\vspace*{-8mm}
\noindent 
\makebox[11.5cm][l]{} August 25, 1998 \\


\vspace{2.cm}
\begin{center}
  {\LARGE {\bf  Improved mass constraints in the MSSM from vacuum stability
 }}\\
\vspace{42pt}
Indranil Dasgupta \footnote{e-mail address: dgupta@physics.uc.edu}\\
Ricardo Rademacher \footnote{e-mail address: ricardor@physics.uc.edu}\\
and\\
Peter Suranyi \footnote{e-mail address: suranyi@physics.uc.edu}\\

\vspace*{0.5cm}

 \ \ Department of Physics, University of Cincinnati \\
{400 Geology/Physics Building, Cincinnati, OH 45221, USA}

\vskip 3.4cm
\end{center}
\baselineskip=18pt

\begin{abstract}
{\normalsize {Using a simple numerical method, we compute the stability of 
the MSSM vacuum with respect to tunneling. The stability criterion is then 
used to put restrictions on the mass parameters. These restrictions are 
necessary conditions for the vacuum stability and complement the existing
sufficiency conditions obtained analytically.

\vspace {0.5 cm}
PACS: 73.40.Gk, 12.60.Jv, 14.80.Ly, 02.70.-c

\vspace {0.5 cm}
Keywords: MSSM, Vacuum Instability, Tunneling, CCB, UFB, Bounce. 
 }}

\end {abstract}

\vfill
\end{titlepage}

\baselineskip=18pt  
\pagestyle{plain}
\setcounter{page}{1}


\section {Introduction}

In this paper we improve constraints on the masses
of the scalar excitations of the minimal supersymmetric Standard Model 
(MSSM) (for a review of the MSSM see ref.~\cite {mssm})
by making use of a new method of computing vacuum tunneling
rates in field theories \cite {dg1}.

To begin with, recall that the most general potential for the scalar fields in the MSSM that respects 
the gauge symetries and conserves baryon and lepton numbers is:
\bear
\label {potone}
V &=&  \sum _j \left | {\partial W \over \partial \phi _j} \right | ^2 + V_D \\
\nonumber &+& \sum _{i,j} m^2 _{ij} \phi_i^*\phi_j  + A_Uh_U{\tilde {Q_L}}H_2{\tilde {U_R}} + 
 A_Dh_D{\tilde {Q_L}}H_1{\tilde D_R} \\ \nonumber
&+&  A_Eh_E{\tilde {L_L}}H_1{\tilde {E_R}} + B\mu H_1H_2
\eear
where $W$ is the superpotential 
\be
\label {pottwo}
W = h_UQ_LH_2 U_R + h_D Q_LH_1D_R +  h_E L_LH_1 E_R + \mu H_1H_2
\ee
and $\phi_j$ stands for any scalar field in the theory ($m_{ij}$ are gauge invariant mass terms). 
$H_1$ and $H_2$ are the Higgs superfields, while
$Q_L, U_R$, and $D_R$ stand for up and down type quark superfields.
$L_L$ and $E_R$
are the lepton superfields ($Q_L$ and $L_L$ are weak $SU(2)$ doublets). 
The Yukawa couplings $h_U, h_D, h_E$, as well as the products 
$A_Uh_U, A_Dh_D$, and $A_Eh_E$ are $3\times 3$ matrices in generation space. 
Superpartners are denoted by a
tilde over the corresponding superfield. The last five terms in equation 
(\ref {potone}) 
are the soft supersymmetry breaking terms. The 
expression for the $D$ term is 
\be
\label {potthree}
V_D = {1\over 2} \sum _a g_a^2 \left ( \sum _{\alpha} \phi _{\alpha} ^{\dagger} T^a \phi_{\alpha} \right )
^2
\ee
where $g_a$ ($a=1,2,3$) are the gauge coupling constants and $T^a$ are the generators of the gauge group
algebra in the representation of the scalar fields $\phi_a$ (namely, the squark, slepton and Higgs fields). 

Even at the classical level, the potential in equation 
(\ref {potone}) is a complicated quartic function which has many local minima.
A systematic study reveals 
that for specific ranges of parameters the potential is
unbounded from below (UFB). Several
charge and/or color breaking (CCB) minima of the potential may also 
be present \cite {ccb,clm,komatsu}. 
Later we will recollect some of the constraints
on the parameter space arising from the requirement
that the charge and color preserving minimum be a global
minimum of the potential. This requirement is however too conservative.
Indeed, it is quite 
possible that the vacuum we live in is a metastable false vacuum with a very long life-time. 
There have been some numerical studies of vacuum decay rates in restricted 
sectors of the MSSM \cite {vacdecay, kls}.
However, untill recently no techniques existed for efficiently scanning the parameter space 
for unstable vacua in the full theory because of the large number of fields involved.
Recently a new method for simplifying the computation of vacuum tunneling
rates in field theories with many scalar fields was introduced by one of us \cite {dg1} 
(in this context, see also ref.~\cite {sarid}). The goal of
the present work is to use this method to relax the mass constraints.
The constraints we 
will obtain by computing a lower bound on the tunneling amplitude of the viable MSSM
vacuum will be
``necessary conditions'' and the parameter space ruled out by these constraints is ``necessarily'' ruled out. 

Our numerical method is described in section 2. 
In section 3 we use this method to constrain the allowed ranges of particle masses. 

\section {The Method of Reduction} 

Suppose we wish to find vacuum tunneling amplitudes in a 4- dimensional field theory described by
the (Euclidean) action 
\bear
\label {actionone}
S &=& T + V\\ \nonumber
T &=& \int d^4 x \left [{1 \over 2}\sum _j \left |{\partial \phi_j \over \partial x_{\mu}}\right |^2\right ] 
\\ \nonumber
V &=& \int d^4 x \left [ U(\phi_1,...\phi_n) \right ]
\eear 
where $\phi_j, j=1,2...,n$ are real scalar fields and $U$ is a potential. Let us also suppose that
$U$ has a local minimum at $\phi_j^f$ (the false vacuum) and a global minimum at $\phi_j^t$ (the 
true vacuum). We normalize $U$ so that $U(\phi_j^f)=0$.
Then the probability density for the rate of nucleation of true-vacuum bubbles in the
false vacuum is given by
\be 
\label {tunnel}
{\Gamma \over V} = A \, {\rm {exp}} \left [ - {S[{\bar {\phi_j}}] \over \hbar} \right ] \left [ 1 + {\rm {O}}
(\hbar) \right ]
\ee
where ${\bar {\phi_j}}$ is a saddle point of the action $S$ (known as the bounce) and $A$ is a pre-factor
of dimension $(mass)^4$ \cite {vko,coleman,cc}. Let us summarize salient features of equation (\ref 
{tunnel}):\\
(i) The bounce is a finite action field configuration which is invariant under time-reversal. It 
follows that ${\bar {\phi_j}}(x) \to \phi_j^f$ as $x \to \infty$ and $\partial {\bar {\phi_j}} (x)
/ \partial t\, |_{t=0} \equiv 0$.\\
(ii) Equation (\ref {tunnel}) is the field theoretic version of a generalized  semiclassical 
WKB tunneling formula in quantum mechanics \cite {bbwl}. The field theoretic generalization
involves renormalization of both $S$ and $A$ \cite {cc}.
If we choose a renormalization scheme where the masses and couplings entering in $S$
are the physical masses and runing couplings at some renormalization scale $Q$, then 
$ {S[{\bar {\phi_j}}] \over \hbar} $ and $A$ are both functions of $Q$
in a way such that $\Gamma \over V$  is $Q$ independent. 
For the MSSM, it has been argued that as long as the mass parameters and the inverse
size of the bounce ${\bar {\phi_j}}$ are within one or two orders of magnitude from
the electroweak scale $Q_0$, the value of $ {S[{\bar {\phi_j}}] \over \hbar} $ and
therefore $A$ is relatively insensitive to radiative corrections provided $Q$ is chosen to be of order
of the electroweak scale \cite {clm, kls}. 
In practice, we will choose $Q = Q_0 \sim 200 {\rm {GeV}}$, find the bounce, compute $S$
and make a simple estimation of $A$ which is calculable in principle \cite {cc, wipf}
but is difficult to calculate exactly even in 
very simple field theories. We will write $A$ as $\eta v^4$ where $v$ is the characteristic mass
scale of the theory (in the present case $v \sim Q_0 \sim 200 \, {\rm {GeV}}$) and $\eta $ is a factor
representing our ignorance. Smallness of the 
radiative corrections imply that the error involved in making the  approximation $\eta \sim 1$ is
negligible (except for a possible enhancement described below). The
approximations made above are valid when the scalar masses and the inverse of bounce size
are not off by several orders of magnitude from $Q_0$. 

(iii) The probability of vacuum decay is large if $\Gamma /V \times T_0^4 \sim 1$ where $T_0$ is the 
age of the universe. The condition $\Gamma /V \ge T_0^4$ translates roughly into
\be
\label {gammav}
{S[{\bar {\phi_j}}] \over \hbar} \le 400 \, 
\ee
when $\eta $ is taken to be of order unity. 
Note that close to the surface separating the regions in the parameter
space where $(\Gamma /V \times T_0^4) -1$ changes sign, a $5\%$ decrease in the tunneling rate 
would require $\eta$ to be as large as ${\rm {exp}(20)}$. There is a recent claim that
$\eta $ may be enhanced by a factor of $[{S[{\bar {\phi_j}}] \over \hbar}]^{N/2}$ 
if the bounce breaks a continuous internal symmetry group of the false vacuum with 
$N$ generators \cite {klw}. We however choose to be conservative and do not
apply this correction which can only strengthen the bounds we obtain on the parameter space in the end.
Also note that we consider only a zero temparature quantum tunneling effect. There is a 
corresponding finite temperature effect for which the tunneling rate will be
larger but the available time for transition will be smaller. The present analysis may
be extended along the lines of \cite {strumia} to obtain independent constraints
from the finite temperature case. 

The most important part of the calculation is the determination of $S[{\bar {\phi_j}}]$ for which we use
the following method of reduction (more details can be found in ref.~\cite {dg1}). Recall that 
the bounce is a saddle
point of the action with a single unstable direction.
It satisfies stationarity with respect to scale transformations $x \to \lambda x$, where $\lambda$ a positive number.
In a $d$ dimensional field theory ($d\ge 2$) the condition of stationarity is \cite {derrick}
\be
\label {TV}
(d-2)T+(d)V=0 \, .
\ee
An interesting property of the bounce is that it is the global minimum of the action in the subspace $C$
of all field configurations with the trivial boundary condition that also satisfy equation (\ref {TV}). 
A rigorous upper bound on the bounce action is obtained by taking any field configuration
$\hat {\phi}_j (x)$ having trivial boundary conditions 
for which $V < 0$ and scale transforming it to satisfy equation (\ref {TV}). A better upper 
bound is obtained by minimizing the action on a finite dimensional subspace of $C$. A good 
numerical algorithm should start with a suitably chosen finite dimensional space over which the minimization
is a quick procedure and pays off well in lowering the action. 
This is the objective of our computations, namely, to obtain
the best possible lower bound on the tunneling rate (given time and computer limitations) and use it to rule out
as much of the parameter space as possible. The method we use is possibly the simplest one
to implement. We remark that a better bound may be obtained by a more computationally
intensive minimization. 

The method of reduction 
is simply a parametric reduction of the number of real fields in the Lagrangian of theory to $1$. For
instance one can define $\phi_j = \alpha_j \Phi, j=1, 2, ..., n$ with
$\sum \alpha_j^2 =1$. This reduces a Lagrangian of $n$ real fields to a Lagrangian of a single real field $\Phi$.
The advantage is an enormous simplification in the numerical search for the minimum of the action 
on $C$. Indeed, with spherical symmetry $\Phi(x) \equiv \Phi (R), R^2 = x_1^2 + ... x_{d}^2$,
the minimization is equivalent to solving the equation of motion:
\be
\label {eqmotion}
{d^2 \Phi \over d R^2} = {dU \over d\Phi} - {(d-1) \over R} {d\Phi \over dR} 
\ee
with the boundary conditions $d\Phi/dR =0$ at $R=0$ and $\Phi \to \Phi^f$ as $R \to \infty$. There is a quick numerical
method called the shooting method for solving this. One starts with a guess for the initial point 
$\Phi (0)$ and numerically
integrates till either $d\Phi/dR$ changes sign (undershoot) or $\Phi$ attains the value $\Phi ^f$ (overshoot).
The initial point for the bounce must lie between the initial point of an overshoot and an undershoot, so the
search converges rapidly by bisecting the interval between two initial point guesses of different outcomes.

The following points in the method of reduction merit a special mention. \\
(i) When some fields are put equal to zero ($\alpha_i =0$) 
some coupling constants may drop out of the Lagrangian. The corresponding
bounds on the parameter space are independent of these couplings.
Thus one can explore particular sets of parameters in the MSSM independently of the 
choice of other parameters. \\
(ii) In general the potential $U$ gives a good indication of the direction in the
field space where the unstable vacuum is most likely to tunnel to. Therefore one can choose the reduction
so that the reduced field $\Phi$ points in the most obviously ``dangerous'' direction.

\section {The Unbounded From Below Directions in MSSM}

Tunneling to a generic CCB vacuum in the MSSM is difficult to explore using the
formalism of section 2 because of the presence of too many masses and couplings. 
We have chosen the directions in the field space where the potential may be 
unbounded from below (UFB) to compute the parameter bounds. In these directions, only
a handful of the scalar fields are important and the reduced action contains 
fewer parameters.

A general classification of UFB directions in the MSSM can be found in ref.~\cite{clm}. 
The unboundedness occurs along directions where the stabilizing 
$D$ and $F$ terms vanish and the soft terms are negative. This can happen in two 
different ways:\\
(I) If $H_1, H_2$ and ${\tilde {L_i}}$ are non-zero and all other scalar fields are zero. The
index $i$ is a generation index here.\\
(II) If either \\
(II a) ${\tilde {E}}_i, {\tilde {L_i}}, {\tilde {L_j}}$ and $H_2$ are non-zero and all
other scalar fields are zero, or \\
(II b) If ${\tilde {D}}_i, {\tilde {Q_i}}, {\tilde {L_j}}$ and $H_2$ are non-zero and all
other scalar fields are zero.

The situations  (II a) and (II b) lead to similar reduced potentials and will be treated
together. Note that the effective potential is unbounded from below due to the tree 
approximation. The complete effective potential is bounded from below. However, when
the quantum corrections to the tree order potential are small, tunneling amplitude to the 
deep CCB vacuum that replaces the UFB direction in the full effective potential should
be calculable from the bounce computed using the tree order potential. This is because
the bounce action actually depends on the ``height'' and ``width'' of the 
potential wall surrounding a false vacuum and is insensitive to the depth of the true vacuum.

Let us write down the potential one obtains by setting all scalar fields
equal to zero except the neutral Higgs fields $H_1^0, H_2^0$ (where we write  
$H_1 = (H_1^0, H_1^{-}), H_2 = (H_2^{+}, H_2^{0})$). A viable MSSM vacuum is obtained
by minimizing this potential.
\be
U(H_1,H_2)=(m_{H_1}^2 + |\mu |^2) H_1^2 + (m_{H_2}^2 + \mu^2) H_2^2 - 2m_3^2 H_1H_2 
+ {1\over 8} [g_1^2 + g_2^2] \left [ H_2^2 -H_1^2 \right ]^2
\label {vh}
\ee
where $ H_i=|H_i^0|, m_3^2= |B\mu|$
and $ g_1$ and $g_2$ are the $U(1)_Y$ and $SU(2)_W$ gauge couplings respectively. We have chosen the phases
of the fields to minimize the potential. Defining $m_i^2 = m_{H_i}^2 + |\mu|^2, (i=1,2),$ we obtain the following 
parametric expressions for the vacuum
expectation values (VEV's) $v_1$ and $v_2$  for $H_1$ and $H_2$ respectively.
\bear
{v_1 \over v_2} &=& {\rm {tan}}\beta\\ \nonumber
v_2^2 &=& {2(m_3^2{\rm {tan}}^2\beta -m_2^2) \over g^2 (1-{\rm {tan}}^2\beta)} 
\label {vonevtwo}
\eear
where $g^2 = g_1^2 + g_2^2$. 
This potential may have a local minimum at $H_1=v_1, H_2=v_2$ with $v_1, v_2 \ne 0$ if
\bear
\label {higgsone}
m_1^2 + m_2^2 & > & 2 m_3^2  \\ \nonumber
m_1^2m_2^2 & < & m_3^4 \, . 
\eear
Either $m_1^2$ or $m_2^2$ can be negative. To get the correct $W$ and $Z$ masses we need
\be
\label {vtwo}
v_1^2 + v_2^2 \approx  (175 {\rm {GeV}})^2
\ee
which fixes the value of $m_3^2$ to be
\be
\label {mthree}
m_3^2 = {|(m_1^2 + m_2^2)| {\sqrt {(m_1^2 + a)(m_2^2 +a)}} \over \left | [m_1^2 + m_2^2 + 2a] \right |} \, 
\ee
with $a \approx {g^2 \over 2} \times (175 {\rm {GeV}})^2$.

\subsection {Case I.}

Writing ${\tilde {L}}_i = ({\tilde 
{\nu}}_L, {\tilde {E}}_L)_i$, the $F$ terms are zero if $H_1^0, H_2^0$ and ${\tilde
{\nu}}_L$ are non-zero but all other scalar fields are zero. The relevant scalar 
potential is:
\be 
\label {va}
U(H,\nu) = {m_1}^2 H_1^2 + m_2^2 H_2^2 - 2m_3^2 H_1H_2 + m^2_{\nu}
{\nu}^2 + {1\over 8} g^2 \left [ H_2^2 -H_1^2 -\nu ^2 \right ]^2
\ee
where $\nu =|{\tilde {\nu}}_{Li}|$. Once again we have chosen the phases to minimize
the potential. Notice that there are no $F$ terms other than those that are already
in equation (\ref {vh}). The $D$ term vanishes identically along the surface $H_2^2-H_1^2-\nu^2 =0$. 
If we define $y= H_1 / H_2$, then the potential on this surface is of the form
\bear
U(H_2)&=&f(y)H_2^2 \\ \nonumber
f(y)&=& \left [ (m_2^2 + m_{\nu}^2) -2m_3^2y + (m_1^2 - m_{\nu}^2)y^2 \right ] \, .
\label {vahtwo}
\eear
The potential is unbounded from below if $f(y)$ is negative. Minimizing $f(y)$ we find the 
direction of fastest decreasing potential along $y= y_m \equiv {m_3^2 \over (m_1^2 -m_{\nu}^2)}$ provided
\be 
\label {conaone}
m_1^2 - m_{\nu}^2 > 0 \, .
\ee
To have a non-zero value of $\nu ^2$ and $H_1^2$ we must have 
\be
\label {conatwo} 
0 <y_m < 1 \, .
\ee
Finally, to have the unboundedness from below we require that
\be
f(y_m) =  (m_2^2 + m_{\nu}^2) - {m_3^4 \over (m_1^2 -m_{\nu}^2)} < 0 \, .
\label {conathree}
\ee
{\centerline {{\epsfxsize=3.0in\epsfbox{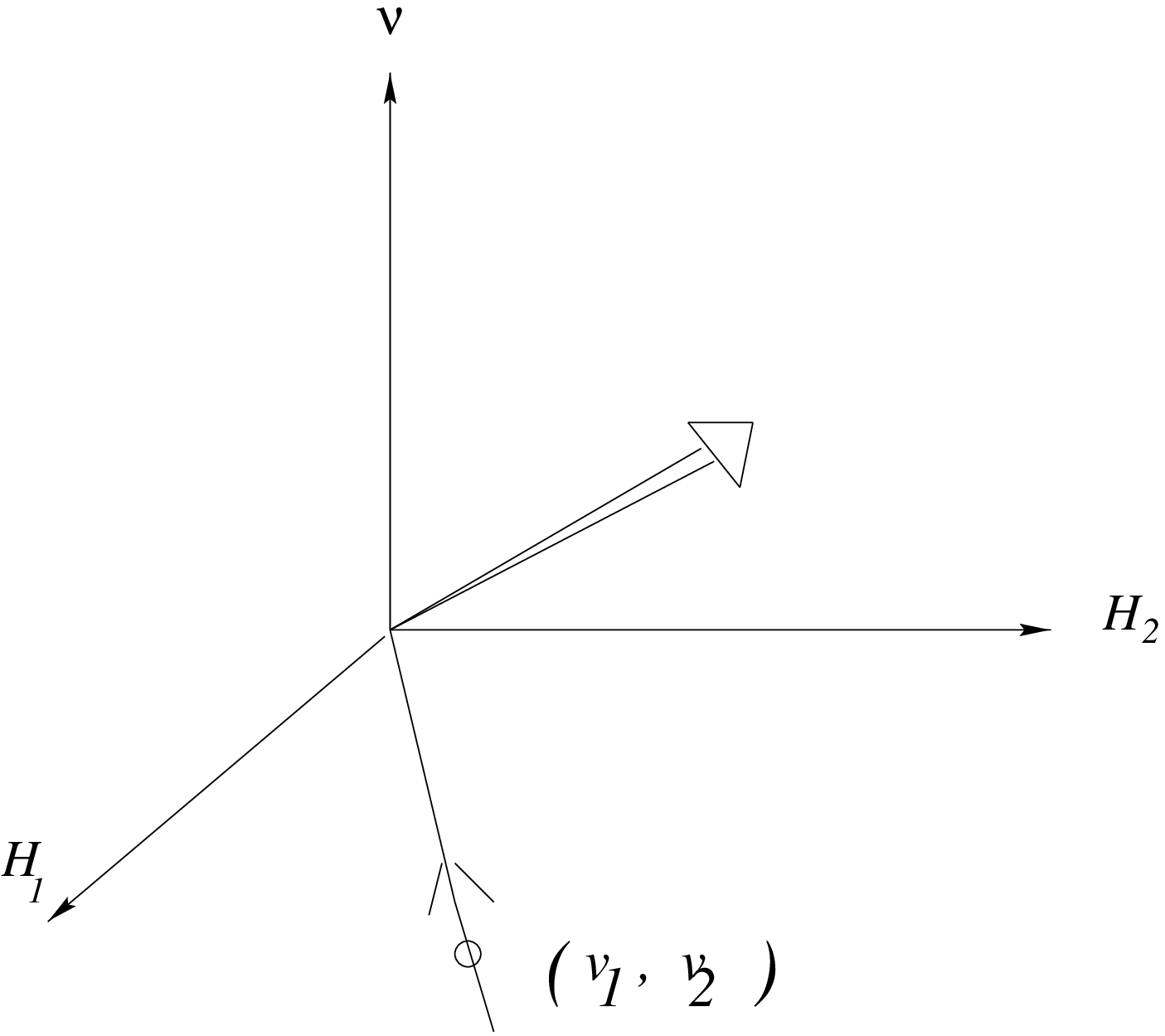}}}}
{\centerline {\makebox[0.8in][l]{\hspace{2ex} Fig. 1}
\parbox[t]{4.8in}{ {\small The reduced field. The piecewise straight line marked with arrows
in the $H_1, H_2, \nu$ plane defines a single real field. The line passes through the vacuum
point $H_1= v_1, H_2 =v_2, \nu =0$. 
}}}}
\vspace {0.5 cm}

The last equation is the sufficiency constraint that we wish to relax to a necessary one.
To compute the decay width of the viable vacuum we reduce the field space as follows. We take a
piecewise straight line in the 3-dimensional space of $H_1, H_2$ and $\nu$ parametrized by the
real number $\phi \in [-\infty, +\infty]$. In the interval $[-\infty, 0]$, we set $|\phi|^2 = H_1^2 + 
H_2^2$ with the points $H_1, H_2$ lying on the line connecting the point 
$H_1=H_2=\nu = 0$ and the 
(viable) vacuum point $H_1=v_1, H_2=v_2, \nu=0$. In the interval $[0, \infty]$, we define $
|\phi|^2 = H_1^2 + H_2^2 + \nu^2$ with the points $H_1, H_2, \nu$ lying on the line that
passes through $H_1=H_2=\nu = 0$ and 
is defined by the relations $H_1 = y_m H_2, H_2^2-H_1^2-\nu^2 =0, H_1, H_2, \nu \ge 0$ [Fig. 1]. 
Along this piecewise straight line the scalar potential reduces to
\be
U(\phi) = \left \{ \begin {array} {rcl} a\phi^4 + b\phi^2 \; &{\rm {for}}& \phi \le 0 \\ [3mm]
{1 \over 2} f(y_m)^2 \phi^2 \; &{\rm {for}}& \phi > 0 \; .
\end {array} \right .
\label {vaphi}
\ee

With this reduction the problem is reduced to solving a differential equation 
for a single real scalar. Note that the kinetic term for the field $\phi$ is
$T = \int d^4x \sum_i({\partial \phi \over \partial x_i})^2$ which reduces to 
$2 \pi^2 \int RdR ({\partial \phi \over \partial R})^2$ upon imposing $O(4)$ symmetry. 
We solve the differential equation for
the bounce, on a computer,
using the shooting method. 
Because the trilinear
and Yukawa couplings do not enter any of the expressions in this case, the results obtained are valid
for all three generations. 

\subsection {Case II.}

Let us develop the formalism by considering the case (II a). 
In this case we have $H_2, E_{Ri}, E_{Li}, \nu_{j} \ne 0, (i\ne j),$ and all other scalar fields are vanishing. We have
retained the tilde-free notation of the previous case and set ${\tilde {E_i}} \equiv E_{Ri}, 
{\tilde {E}}_{Li} \equiv E_{Li}$. After setting $F =
|\mu H_2 + h_{Ei} E_{Li}E_{Ri}|^2 =0$, and $|E_{Ri}|=|E_{Li}| = E$, one gets the potential:
\be
\label {vb}
U(H_2,E,\nu) = (m_2^2 - |\mu|^2) H_2^2 + (m^2_{Li} + m^2_{Ei})E^2 + m^2_{Lj}\nu^2 + {1 \over 8}g^2
\left [H_2^2 + E^2 - \nu^2 \right ] \, ,
\ee
where we have simplified notation with $|\nu_j| \to \nu$. 
Note that $h_{Ei}$ is just a number.
The $D$ term is zero if $\nu^2 = H_2^2 + E^2$. The potential in 
this direction is:
\be
\label {vbhtwo} 
U^{\prime}(H_2) = (m_2^2 -|\mu|^2 + m_{Lj}^2)  H_2^2  \pm {|\mu| \over h_{Ei}} (m_{Li}^2 + m_{Ei}^2 + 
m_{Lj}^2) H_2 \, .
\ee
The ambiguous sign comes from phase choices of the fields. We will choose the sign that makes the
second term on the right negative (\ie a faster drop in the potential). 
The resulting 
sufficiency condition for vacuum stabilty is a strong constraint (the Komatsu constraint \cite {komatsu}):
\be
m_2^2 -|\mu|^2 + m_{Lj}^2 \ge 0 \, .
\label {komatsu}
\ee
The field reduction 
is now slightly more complicated than that of the previous case. 
We define a real scalar $\phi \in [-\infty, +\infty]$ that for $\phi > 0$ is the path length
of the curved line defined by the equations $H_2^2+E^2-\nu^2 =0$ and $F=0$ measured from the
point $H_2=\nu=E=0$. When $\phi <0$ it is given by $\phi^2 = H_1^2 + H_2^2$ with $H_1=y_mH_2$, as
in the previous section. The infinite (curved) line $\phi \in [-\infty, +\infty]$ lies in the 5- dimensional
Euclidean space with coordinates $H_1, H_2, E_{Li}, E_{Ri},
\nu$. The scalar potential along this curved line is
\be
U(\phi ) = \left \{ \begin {array} {rcl}  a\phi^4 + b\phi^2 \, &{\rm {for }}& \, \phi < 0 \\ [3mm]
U^{\prime}(H_2) \, &{\rm {for }}& \, \phi \geq 0 \, . \end {array} \right . 
\label {vbphi}
\ee
The parametric representation of the potential in terms of $H_2$ hides the main complexity of this 
case, namely, $\phi$ (when positive) must be computed as a function of $H_2$ by integrating a differential
along the curve defined above:
\be
\label {phidiff}
d\phi = \sqrt {\left [ dH_2^2 + dE_{Li}^2 + dE_{Ri}^2 + d\nu^2 \right ] } \, .
\ee
The case II b is trivially obtained by the substitution $E_i \to D_i, 
L_i \to Q_i$ and $h_{Ei} \to h_{Di}$ in equations (\ref {vb}) and (\ref {vbhtwo}).

\section {Results}

\subsection {Case I.}

\bigskip
{\centerline {{\epsfxsize=3.0in\epsfbox{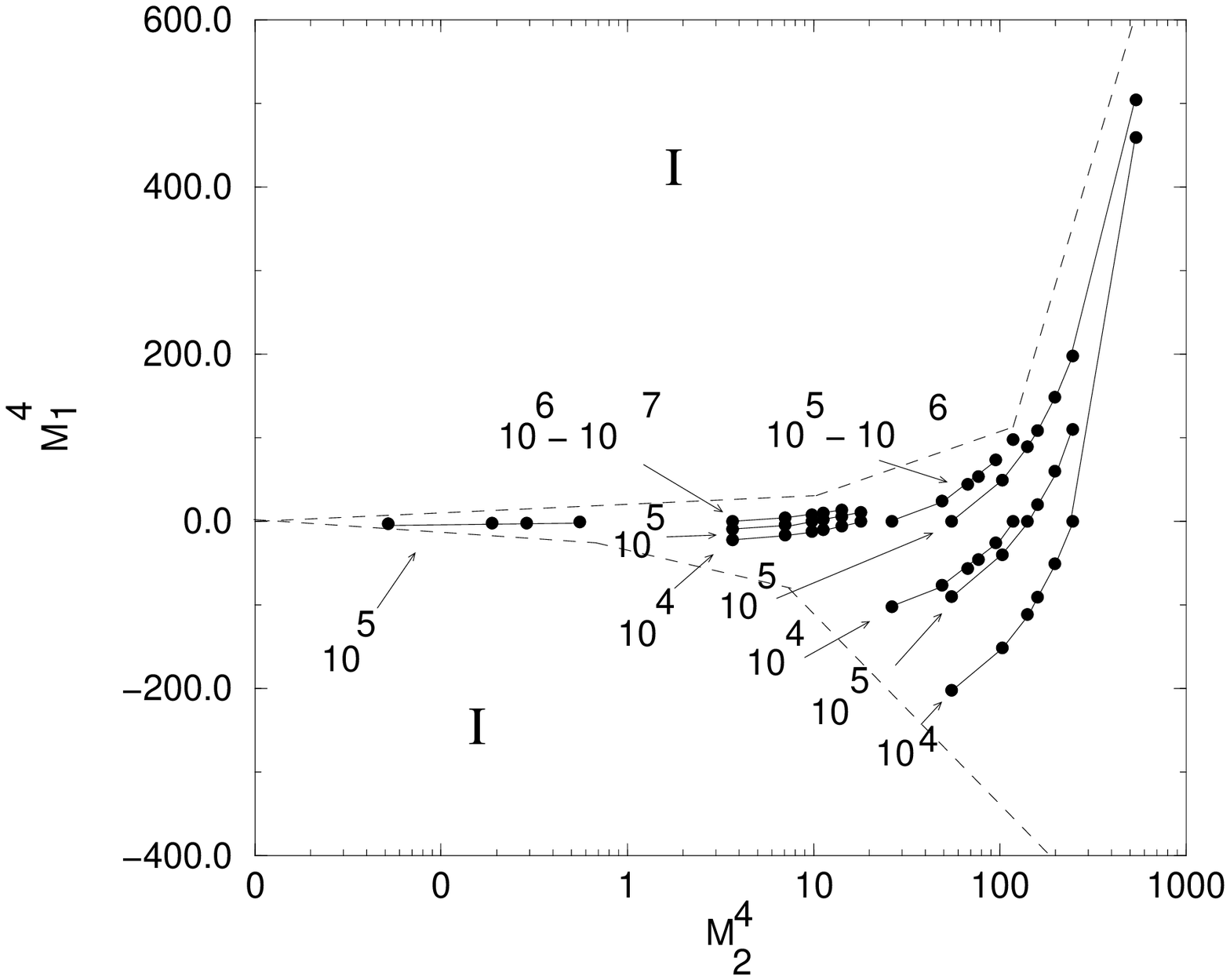}}}}
\makebox[0.8in][l]{\hspace{2ex} Fig. 2.}
\parbox[t]{4.8in}{ {\small Bounce actions in Case I. We have drawn iso-action lines connecting
points with action (shown adjacent to lines) of same order. All masses are in units of $100 \, {\rm GeV}$. 
The regions marked ``I'' on either side of broken lines have no points relevant to tunneling computation.
}}
\vspace {0.5cm}

In this case the requirement that a viable MSSM vacuum exist
(equations (\ref {higgsone}, \ref {mthree})) is hard to satisfy 
simultaneously with the requirement that destabilizing directions exist 
(equations (\ref {conaone}, \ref {conatwo}, \ref  {conathree})). It is convenient to plot the results 
in the $M_1^4,
M_2^4$ plane where we define $M_1^4 = (m_2^2 + m_{\nu}^2) (m_1^2 -m_{\nu}^2)$ 
and  $M_2^4 = m_3^4$. 
The compression of the 4- dimensional 
parameter space ($m_1, m_2, \mu, m_{Li}$) to this plane leads to little loss of information
since the bounce action is relatively insensitive to the other two axes. In Fig.~2 regions
marked with an ``I'' 
do not satisfy the requirements mentioned above. This leaves a narrow
window in the realistic mass range open for our computations where the bounce actions
are greater than $400 \hbar$ (usually by an order of magnitude). We have drawn iso-action lines
for the bounces to display the action variation. Since we are unable to rule out 
any part of the parameter space with certainty, this case serves to display the limitation
of our algorithm. However the plot shows the range and
variation of the bounce action and the result here may be improved by a more accurate numerical
calculation. 

\subsection {Case II a.}

We have reduced the number of free parameters by using approximate degeneracy of the slepton masses.
The restriction to this subspace of the parameter space is motivated by Grand Unification 
of couplings and masses (for a review see ref.~\cite {mssm}). To be conservative, we chose the case of the
$\tau $ slepton (\ie $i=3$ in the relevant formulas) 
which has the largest Yukawa coupling and (given slepton mass unification) 
from inspection of equation (\ref {vbhtwo}),
is likely to give the least negative contribution to $U^{\prime}(H_2)$. 
The results are not very sensitive to slepton mass splittings. Again
it is convenient to define the masses $ M_3^2 = - (m_2^2 - |\mu |^2 + m_{Lj}^2)$, and $M_4^3 =
{3 \over h_{E}} |\mu | m_{Lj}^2$. The plot of bounce actions on the $M_3^2$ vs. $M_4^3$ plane is relatively
insensitive to other independent axes and shows a large region where our methods are adequate
to prove the instability of the MSSM vacuum (Fig.~3). 
We have scanned more than $2 \times 10^5$ points in the the parameter region 
$m_1^2 \in [-100, 500], m_2^2 \in [-2, 100], \mu^2 \in [0, 500]$ and $m_{Lj}^2 \in [0.2, 100]$
in units of $(100 \, {\rm {GeV}})^2$. A fraction of these points appear in Fig.~3 which 
contains our result in the form of the broken line. 
{\it {To the right of the line a stable and viable MSSM vacuum is ruled
out}} because at least one of the following necessarily happens:
(i) equations (\ref {higgsone}), (\ref {mthree}) are not satisfied,
(ii) slepton mass $< 45 \, {\rm GeV}$, (iii) ${S\over \hbar }  < 400$. 
{\it {To the left of the line a stable and viable MSSM vacuum may or may
not exist}}. To obtain this line, one need only plot 
those points for which equations (\ref {higgsone},),
(\ref {mthree}) {\it {are}} satisfied and slepton mass $< 45 \, {\rm GeV}$ and then draw the line
to bound all points with ${S\over \hbar } > 400$ (the crosses) from the right. Indeed, the points with 
${S\over \hbar }  < 400$ need not be plotted either. In Fig.~3, 
we plot a fraction of the points that have  ${S\over \hbar }  < 400$ to illustrate
our coverage of the parameter space.

The broken line is therefore our main result. It is a modification of the Komatsu constraint to the 
approximate relation for vacuum instability:
\be 
M_3^2 x + 0.2 M_4^3 \ge 9.5 x^3 \, 
\ee
valid for $M_3 \ge 0$ with $x = 100 \, {\rm GeV}$ and when all scalar masses in the theory are
within two orders of magnitude of the electroweak scale. 
The simple method used by us is successful in 
ruling out most of the relevant region in the parameter space.

{\centerline {{\epsfxsize=4.0in \epsfbox{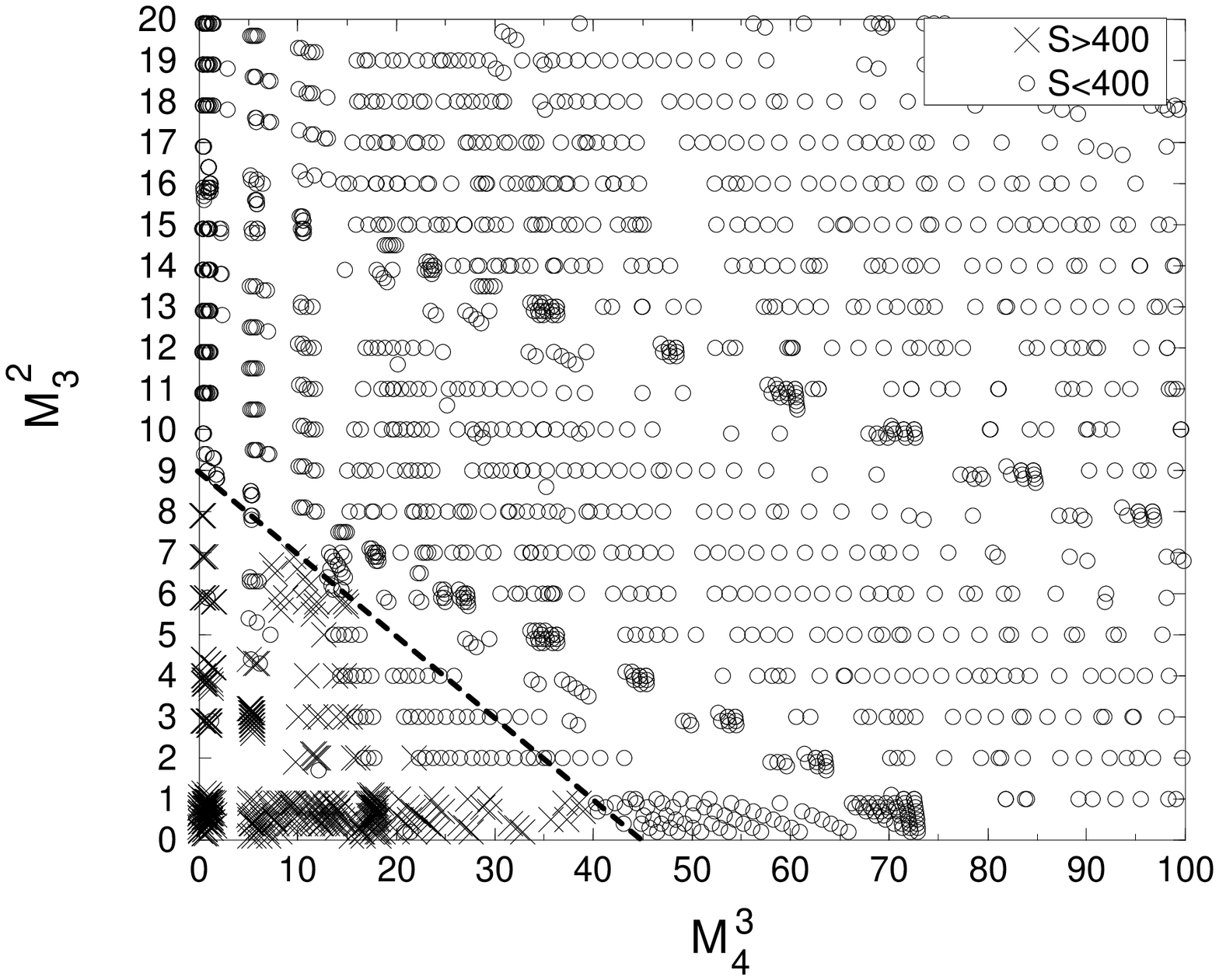}}}}
\makebox[0.8in][l]{\hspace{2ex} Fig. 3.}
\parbox[t]{4.8in}{ {\small Bounce actions in Case II a. 
Crosses denote points with $S/\hbar > 400 $. The circles mark points with $S/\hbar < 400 $. 
The action tends to decrease as one moves to the right or upwards of the broken line
and only points with $S/\hbar < 400 $ appear outside the window we display.
All masses are in units of $100 \, {\rm GeV}$. }}
\vspace {0.5cm}

\subsection {Case II b.}

This case is identical to the previous one with the replacement ${L_i} \to {Q_i}, E_i \to D_i$ and 
$h_{Ei} \to h _{Di}$. Again we use approximate degeneracy of squark and
slepton masses and study the bottom squark which has the largest Yukawa coupling  
($i=3$). We define  $ M_5^2 = -(m_2^2 - |\mu |^2 + m_{Qj}^2)$, and $M_6^3 =
{3 \over h_D} |\mu | m_{Lj}^2$. Scanning the same parameter region as in the previous
case we obtain a plot of bounce actions on the $M_5^2$ vs. $M_6^3$ plane (Fig.~4) 
which is  similar to the plot in the previous case and rules out a comparable
volume in the parameter space. The corresponding modification of the Komatsu constraint is the
following relation for vacuum instability:
\be
M_5^2 x + 0.15 M_4^3 \ge 6.0 x^3 \, 
\ee
valid for $M_5 \ge 0$, with $x = 100 \, {\rm GeV}$. 

{\centerline {\epsfxsize=4.0in {\epsfbox{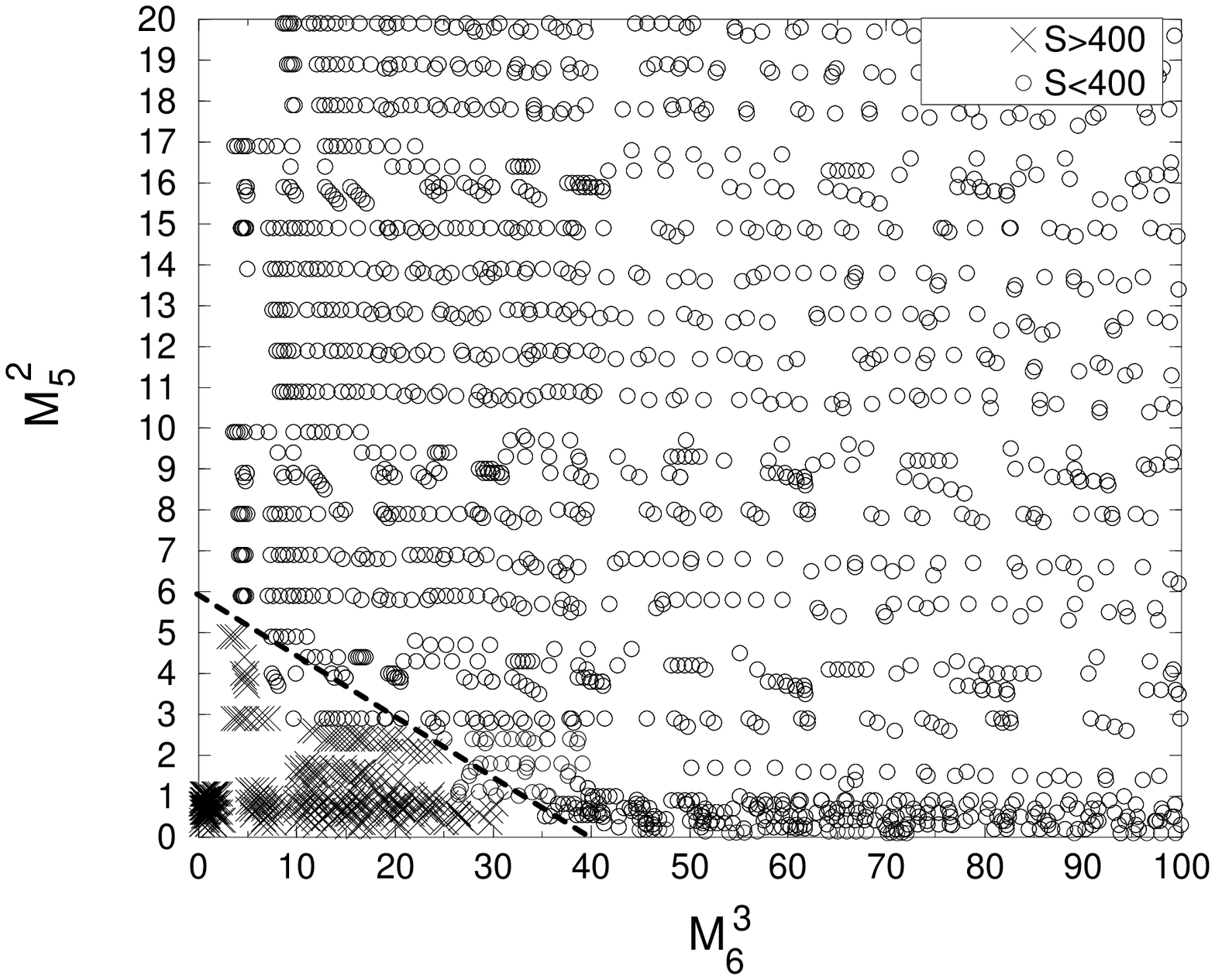}}}}
{\centerline {\makebox[0.8in][l]{\hspace{2ex} Fig. 4.}
\parbox[t]{4.8in}{ {\small Bounce actions in Case II b. The conventions of Fig. 3 apply.
}}}}
\vspace {0.5cm}

As mentioned in section 3, the action values are rigorous upper bounds on the bounce action provided
one rescales the bounce points to lie on the space $C$. The rescaling factor departs
from $1$ by less than $5$ percent ($[T+2V]^2/T^2$ is less
than $10^{-3}$ in all cases). We also found the bounce size to be between $1$ and $10$ in units of $(100 \,
{\rm {GeV}})^{-1}$ which justifies the approximation $\eta = 1$.

\section {Conclusions}

In conclusion, we have used a new and simple method of computing vacuum decay rates in 
field theories to scan the parameter space of the MSSM and test for vacuum stability. We
find that in one class of the characteristic UFB direction, our method fails to yield a useful constraint.
However it does successfully compute a lower bound on the decay rate. In another class of the UFB
direction our method yields very strong constraints and we are successful in ruling out 
with certainty a significant part of the parameter space where the MSSM vacuum is metastable.

{\centerline {\bf {Acknowledgements}}}

We would like to thank Dru Renner for his help in the computational
part of this project. This work was supported by the Department of Energy under the grant
DE-FG02-84ER40153.

\vfill


\vspace{1.0cm}

\begin{thebibliography}{99}

\bibitem {mssm} H.~P.~Nilles, Phys.~Rep.~{\bf 110}, 1 (1984).

\bibitem {dg1} I.~Dasgupta, Phys.~Lett.~{\bf B 394}, 116 (1997), BUHEP-96-31, hep-ph/9610403.


\bibitem {ccb} J.~M.~Frere, D.~R.~T.~Jones and S.~Raby,  Nucl.~Phys.~{\bf B 222}, 11 (1983);
L.~Alvarez-Gaume, J.~Polchinski and M.~Wise,  Nucl.~Phys.~{\bf B 221}, 495 (1983);
J.~P.~Derendinger and C.~A.~Savoy,  Nucl.~Phys.~{\bf B 237}, 307 (1984); 
C.~Kounnas, A.~B.~Lahanas, D.~V.~Nanopoulos and M.~Quiros,  Nucl.~Phys.~{\bf B 236}, 438 (1984);
M.~Drees, M~Gl\"{u}ck and K.~Grassie, Phys.~Lett.~{\bf B 157}, 164 (1985);
J.~F.~Gunion, H.~E.~Haber and M.~Sher,  Nucl.~Phys.~{\bf B 306}, 1 (1988).

\bibitem {clm} J.~A.~Casas, A.~Lleyda and  C.~Munoz, 
Phys.~Lett.~{\bf B 389}, 305 (1996), hep-ph/9606212.

\bibitem {komatsu} H.~Komatsu, Phys.~Lett.~{\bf B 215}, 323 (1988). 

\bibitem {vacdecay} M.~Claudson, L.~J.~Hall and I.~Hinchliffe, Nucl.~Phys.~{\bf B 228}, 501 (1983).

\bibitem {kls} A.~Kusenko, P.~Langacker and G.~Segre,  Phys.~Rev.~{\bf D 54}, 5824 (1996), 
hep-ph/9602414. 

\bibitem {sarid} U.~Sarid, hep-ph/9804308.

\bibitem {vko} M.~B.~Voloshin, I.~Yu.~Kobzarev and L.~B.~Okun', Yad.~Fiz.~{\bf 20}, 1229 (1974);
Sov.~J.~Nucl.~Phys.~{\bf 20}, 644 (1975).

\bibitem{coleman}  S.~Coleman, Phys.~Rev.~{\bf D 15}, 2929 (1977).

\bibitem {cc} C.~G.~Callan and S.~Coleman, Phys.~Rev.~{\bf D 16}, 1762 (1977).

\bibitem {bbwl} T.~Banks, C.~Bender, T.~T.~Wu, Phys.~Rev.~{\bf D 8}, 3366 (1973);
J.~S.~Langer, Ann.~Phys.~(N.Y)~{\bf 41}, 108 (1967).

\bibitem {wipf} A.~W.~Wipf, Nucl.~Phys.~{\bf B 269}, 24 91986).

\bibitem {klw} A.~Kusenko, K.~Lee, E.~J.~Weinberg, Phys.~Rev.~{\bf D 55}, 4903 (1997), hep-th/9609100.

\bibitem {strumia} A.~Strumia, Nucl.~Phys.~{\bf B 482}, 24 (1996), hep-ph/9604417.

\bibitem {derrick} G.~H.~Derrick, J.~Math.~Phys.~{\bf 5}, 1252 (1964).













\end{thebibliography}
\end{document}